\documentclass[sigconf]{acmart}

\usepackage{graphicx}
\usepackage{listings} 
\usepackage{cite}
\usepackage{booktabs}
\usepackage{multirow}
\usepackage{adjustbox}
\usepackage{hyperref} 

\AtBeginDocument{%
  \providecommand\BibTeX{{%
    \normalfont B\kern-0.5em{\scshape i\kern-0.25em b}\kern-0.8em\TeX}}}

\copyrightyear{2021}
\acmYear{2021}
\setcopyright{ccby}
\acmConference[K-CAP '21]{Proceedings of the 11th Knowledge Capture Conference}{December 2--3, 2021}{Virtual Event, USA}
\acmBooktitle{Proceedings of the 11th Knowledge Capture Conference (K-CAP '21), December 2--3, 2021, Virtual Event, USA}
\acmDOI{10.1145/3460210.3493561}
\acmISBN{978-1-4503-8457-5/21/12}

\hypersetup{draft}

\begin{document}
\fancyhead{}
\title{Expressing High-Level Scientific Claims with Formal Semantics}

\author{Cristina-Iulia Bucur}
\affiliation{
	\institution{Vrije Universiteit Amsterdam}
	\city{Amsterdam}
	\country{The Netherlands}}
\email{c.i.bucur@vu.nl}

\author{Tobias Kuhn}
\affiliation{
	\institution{Vrije Universiteit Amsterdam}
	\city{Amsterdam}
	\country{The Netherlands}}
\email{t.kuhn@vu.nl}

\author{Davide Ceolin}
\affiliation{
	\institution{Centrum Wiskunde \& Informatica}
	\city{Amsterdam}
	\country{The Netherlands}}
\email{davide.ceolin@cwi.nl}

\author{Jacco van Ossenbruggen}
\affiliation{
	\institution{Vrije Universiteit Amsterdam}
	\city{Amsterdam}
	\country{The Netherlands}}
\email{jacco.van.ossenbruggen@vu.nl}

\renewcommand{\shortauthors}{Bucur, et al.}

\begin{abstract}
The use of semantic technologies is gaining significant traction in science communication with a wide array of applications in disciplines including the life sciences, computer science, and the social sciences. Languages like RDF, OWL, and other formalisms based on formal logic are applied to make scientific knowledge accessible not only to human readers but also to automated systems. These approaches have mostly focused on the structure of scientific publications themselves, on the used scientific methods and equipment, or on the structure of the used datasets. The core claims or hypotheses of scientific work have only been covered in a shallow manner, such as by linking mentioned entities to established identifiers. In this research, we therefore want to find out whether we can use existing semantic formalisms to fully express the content of high-level scientific claims using formal semantics in a systematic way. Analyzing the main claims from a sample of scientific articles from all disciplines, we find that their semantics are more complex than what a straight-forward application of formalisms like RDF or OWL account for, but we managed to elicit a clear semantic pattern which we call the ``super-pattern''. We show here how the instantiation of the five slots of this super-pattern leads to a strictly defined statement in higher-order logic. We successfully applied this super-pattern to an enlarged sample of scientific claims. We show that knowledge representation experts, when instructed to independently instantiate the super-pattern with given scientific claims, show a high degree of consistency and convergence given the complexity of the task and the subject. These results therefore open the door on the longer run for allowing researchers to express their high-level scientific findings in a manner they can be automatically interpreted. This in turn will allow for automated consistency checking, question answering, aggregation, and much more.
\end{abstract}



\maketitle

\section{Introduction}
\label{sec:introduction}

Semantic Web technologies have started to be more widely applied in science communication to address, among other things, the accelerating growth of scientific literature \citep{Shotton2009_2}. This growth makes it increasingly difficult for individual researchers to follow and be up-to-date with all the current developments in their fields \citep{Landhuis2016}. However, despite semantic technologies being applied in many ways in science, there still is a big gap between the existing (and continuously developing) formal ontologies and the content of scientific articles expressed in natural language that is only human-readable \citep{Shotton2009_1}. As such, the content of scientific articles of today is not accessible to algorithms. Methods of semantic interlinking and metadata enrichment of scientific articles exist, especially in life sciences \citep{Ciccarese2011, GarciaCastro2013, Ciccarese2012, Sernadela2017}, but they are mostly about annotating the existing natural language text in a relatively shallow manner. Other successful applications of semantic technologies focus only on the metadata level, such as provenance representation and versioning of scientific works \citep{Ciccarese2011}.

Applying semantic technologies to not just annotate or describe (on a metalevel) but actually express scientific knowledge is a much more ambitious and much harder task. While technologies like RDF and OWL grounded in first-order logic are now mature and well-tested, full semantic representation of natural language has remained a task that is too complex for automated approaches.
Even the extraction of simpler RDF-based structures from the content of scientific articles typically requires manual curation to achieve sufficient level of quality \citep{GarciaCastro2013, Coulet2011, Sernadela2015}.
In fact, even for knowledge representation experts it is very challenging to fully represent typical high-level scientific findings in formal logic, as no general scheme or procedure exists that is known to apply to findings across specific fields.

In order to make progress on these challenges, we focus here on how general high-level scientific claims can be represented in formal logic. We assume in this work that these representations are manually created, but it could also lay the basis for future research on how to automatically generate these representations from scientific texts (though we assume this to be a very hard problem, as it has been in the past).
Concretely, we introduce below our approach to express the content of high-level scientific claims with formal semantics with a semantic template that we call the ``super-pattern''.

In this research we therefore aim to answer the following research questions:
\begin{enumerate}
    \item To what extent can our super-pattern be used to formalize the main claims of scientific articles from different disciplines?
    \item How reliably can the super-pattern be applied to formalize existing claims by knowledge representation experts?
\end{enumerate}

We envisage that in the longer-term future researchers could themselves express their findings with the super-pattern and thereby make their work directly add to a complex knowledge graph of scientific findings. On that basis, it is easy to imagine how technologies like graph query languages and logic reasoners can be used to find similar research, corroborate scientific claims, spot contradictions, provide aggregations and visualizations, answer questions, and many other kinds of tasks. Among countless other applications, this could save months of literature study for first-year PhD students to get an overview of their fields.

\section{Background}
\label{sec:background}

Despite being in the digital era, the medium in which the scientific articles are still published are journals, which are now electronically available, but otherwise still follow the old paper paradigm \citep{Peroni2012, Peroni2014, Shotton2009_1}. On top of this, the growing number of articles that are published every day makes it imperative to make the articles machine-readable as well, as human researchers are reaching their limits in terms of how much they are able to process and read every day \citep{Landhuis2016, Roetzel2019}. With current developments in the Semantic Web, with technologies like RDF, OWL, and SPARQL \citep{Shotton2009_1, Shotton2009_2}, it is currently possible to enrich the meaning of a traditional article in the digital publishing environment and facilitate its automatic discovery, to have access in a semantic way within the article and link to other related articles or other related parts of articles \citep{Mirri2017, Sateli2016, Jacob2017}.

Based on these Semantic Web technologies, there are various approaches that try to automatically extract such semantic information from articles to make them more machine-readable, covering the metadata \citep{Ciccarese2012} or the article structure \citep{Taakis2015}, providing summarizations of the main concepts and ideas (human assisted and curated) \citep{Sernadela2015}, using annotations to semantically enhance articles \citep{Ciccarese2011, Taakis2015, Sernadela2017} or semantically interlink them \citep{GarciaCastro2013}, and describing bibliographic references and citations \citep{Peroni2012a}. On the one hand, we have rich vocabularies and ontologies like the SPAR ontology suite \citep{Peroni2008} that support the classical publishing process starting from formal descriptions of the structure of articles and article discourses \citep{Constantin2016}, to the traditional publishing workflows \citep{Gangemi2014}, to accounting for all roles involved in the publishing process \citep{Peroni2012PRO}, and even to the current imperfect peer-reviewing process \citep{FROntology2018}. On the other hand, there are Artificial Intelligence tools that can identify plagiarism and predict the impact an article might have \citep{Razak2021}, to natural language processing techniques that make use of strategies such as entity extraction to identify concepts and ideas in articles, assisted and curated by humans \citep{Sernadela2015}.

Data infrastructures such as Linked Open Data allow for connecting data between scientific publications \citep{Sikos2017}, but this needs additional steps as concepts and relations need to be identified beforehand, and the content of the articles need to be interconnected semantically, usually by using humans in the loop \citep{Ciccarese2012, Sernadela2017}. Moreover, more complex approaches that attempt to automate the process of extracting formal semantics from traditionally-structured scientific articles include techniques like the compositional and iterative semantic enhancement method (CSIE) \citep{Peroni2017_1}, semantic lenses \citep{diIorio2014}, and modelling the context of sentences from scientific articles in conceptual frameworks \citep{Angrosh2014}. Semantic Web technologies are therefore extensively used but the underlying publishing paradigm has stayed the same \citep{Stern2019}.

New initiatives that try to change this old paradigm of publishing are proposed, together with new publishing workflows. This paradigm shift entails to move from the textual representation of information to a more data-centric one, similar to the proposal for the next-generation Web \citep{BernersLee2001}, where instead of documents, the interest shifts from the syntactic (e.g. HTML) to the semantic level (e.g. RDF, OWL). On the semantic level, we would be able to express the content and not just the structure of what is now in narrative documents. Moreover, formats that are based on HTML, like RASH \citep{Peroni2017_2}, have been proposed, where scientific articles that include semantic annotations can be represented. Other initiatives involve using semantic representations from the start, written by the actual authors of the research in what is named \textit{genuine semantic publishing} \citep{Kuhn2017_1}, and we have proposed in our earlier work to move from the monolithic form structure of scientific articles to smaller, more granular interconnected parts that each contain single scientific claims or statements from the start \citep{Bucur2020}.

The Open Research Knowledge Graph \citep{Jaradeh2019} is another initiative that aims to make research articles machine-readable. With this approach, scientific entities are expressed as a semantically interconnected knowledge graph, populated by methods such as extracting scientific concepts from the abstracts of scientific articles with the help of annotators \citep{Brack2020}. In the life sciences, especially in the biomedical fields, there have been many initiatives to create controlled vocabularies that can serve as the foundation to represent scientific knowledge in a structured way in order to capture evidence and scientific findings from research in a computable form \citep{Chibucos2014, Slater2012, Madan2019} with specific markup languages \citep{Hucka2003} and exchange formats \citep{Demir2010}. Some of these approaches target high-level claims, but only with restricted coverage, for example BEL \citep{Slater2014} for specific kinds of biological relations.

Nanopublications \citep{Groth2010} are a concept and technology of Linked Data containers that can be used to represent and share different kinds of scientific knowledge. They are expressed in a way that is fully formal, thus machine-interpretable and can be regarded as ``minimal publications'' due to the fact that they contain just three basic elements (represented in RDF): an assertion containing a small unit of information that is its main content (such as a scientific finding), provenance of the assertion (e.g. linking to the scientific methods used to derive the scientific finding in the assertion) and a publication information part about the nanopublication as a whole (e.g. when it was created and by whom). Nanopublications can not only be an openly accessible structured data container for scientific findings or claims, but they can also be used for different types of meta-level assertions as well, for example for statements or assessments about other nanopublications \citep{kuhn2013broadening}.

\section{Approach and Methods}
\label{sec:approach-methods}

We present here our approach to formalize high-level scientific claims by proposing what we call a ``super-pattern''. This super-pattern is a general template of a logical statement that can be instantiated to represent scientific claims in formal logic. To aid the development of this super-pattern, we created a dataset of 50 claims from scientific publications (Dataset A). After the super-pattern had been finalized, we created a second set of another 25 claims (Dataset B), which did not influence our design decisions of the super-pattern and is therefore not biased towards it.

\subsection{Super-Pattern}
\label{sec:super-pattern}

The super-pattern is a template with five slots, which are to be filled in when it is instantiated. It has the following structure with five elements with expected types in square brackets:
\begin{itemize}
\item \textbf{Context class} (``in the context of all ...''): [class identifier]
\item \textbf{Subject class} (``things of type ...''): [class identifier]
\item \textbf{Qualifier} (e.g. ``mostly''): [qualifier from closed list] 
\item \textbf{Relation type} (``have a relation of type ...''): [relation type from closed list]
\item \textbf{Object class} (``to things of type ... that are in the same context''): [class identifier]
\end{itemize}
The context class is optional, whereas all other slots are mandatory. The phrases in quotes above give some hints towards how to interpret it, but we will provide a thorough formal definition below. But let us first have a look at a concrete example.

\begin{figure}[tbp]
	\centering
	\includegraphics[width=\linewidth]{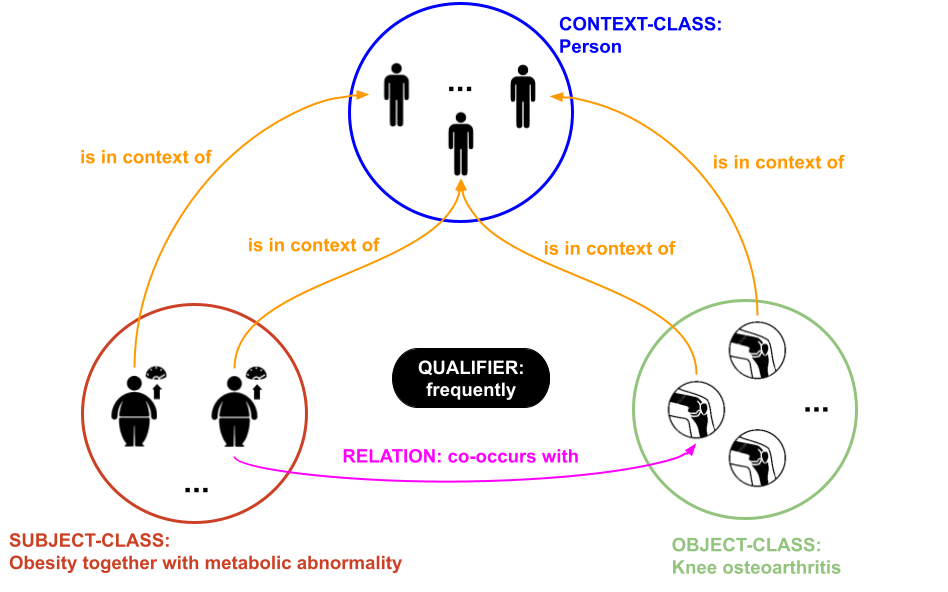}
    \caption{\label{fig:superpattern-diagram}Diagram with an example for the Super-pattern instantiation.}
\end{figure}

In Figure \ref{fig:superpattern-diagram} we see an example where the context class is named ``person'', the subject class is named ``obesity together with metabolic abnormality'', the qualifier is ``frequently'', the relation is ``co-occurs with'' and the object class has the name ``knee osteoarthritis''. By just filling in the informal phrases above, with a little bit of editing we get the following sentence, hinting at how to interpret it:

\begin{quote}
In the context of all \emph{persons}, things of type \emph{obesity together with metabolic abnormality} \emph{frequently} have a relation of type \emph{co-occurs with} to things of type \emph{knee osteoarthritis} that are in the same context (i.e. the same \emph{person}).
\end{quote}
It is easy to see that this is saying that when people have obesity together with metabolic abnormality then these people frequently also have a thing called knee osteoarthritis.

\begin{table}[tbp]
\caption{Super-pattern qualifiers.}
\label{tab:qualifiers}
\centering
\small
\begin{tabular}{l|l|ll}
qualifier & interpretation & $\lesseqgtr$ & $q$ \\
\hline
(can) always & 100\% & $=$ & 1 \\
(can) generally & at least 90\% & $\geq$ & 0.9 \\
(can) mostly & at least 50\% & $\geq$ & 0.5 \\
(can) frequently & at least 10\% & $\geq$ & 0.1 \\
(can) sometimes & at least 0.1\% & $\geq$ & 0.001 \\
\hline
(can) never & 0\% & $=$ & 0 \\
(can) generally not & at most 10\% & $\leq$ & 0.1 \\
(can) mostly not & at most 50\% & $\leq$ & 0.5 \\
(can) frequently not & at most 90\% & $\leq$ & 0.9 \\
(can) sometimes not & at most 99.9\% & $\leq$ & 0.999 \\
\end{tabular}
\end{table}

To arrive at a more formal representation grounded in logic, we can start by looking at the qualifiers like ``generally''. Our super-pattern defines the qualifiers shown in Table \ref{tab:qualifiers}. There are the five base qualifiers \emph{always}, \emph{generally}, \emph{mostly}, \emph{frequently}, and \emph{sometimes}, each with its negative counterpart, such as \emph{never} for \emph{always} and \emph{frequently not} for \emph{frequently}. On top of that, each of these ten qualifiers can be amended by the operator \emph{can}, such as \emph{can generally}, which will be interpreted in a modal logic manner. The table shows how each qualifier is mapped to a quantified interpretation, such as \emph{at most 90\%} for \emph{frequently not}. These interpretations are necessarily a bit arbitrary, but to reach our goal of fully formal semantics we had to define a precise value for these otherwise vague terms.

For our formalization to be described below, each of the 20 qualifiers is given a relation predicate $\lesseqgtr$, a probability value $q$, and a modality value $m$. The values for $\lesseqgtr$ and $q$ are shown in Table \ref{tab:qualifiers}. The modality value $m$ is $\Diamond$ for the qualifiers with \emph{can} and the empty symbol $\epsilon$ otherwise.

We define the semantics of an instantiated super-pattern by the following logic expression template:
\begin{multline*}
P( \mbox{~} m( \exists z( o(z) \wedge i(z,x) \wedge r(y,z) ) ) \mbox{~} | \mbox{~} s(y) \wedge c(x) \wedge i(y,x) \mbox{~} ) \lesseqgtr q
\end{multline*}
Here, $c$, $s$, and $o$ stand for the context, subject, and object classes, respectively, of the instantiated super-pattern. Them being classes they map to unary predicates in the expression above. The super-pattern qualifier is expressed with the three parameters $\lesseqgtr$, $q$, and $m$, as explained above. $r$, finally, represents the relation of the instantiated super-pattern, mapped to a binary predicate. This relation comes from a fixed list of possible relations, consisting of sameness, numerical comparison, causality, and spatio-temporality relations. The full list of relations can be found in the online super-pattern ontology documentation\footnote{\url{https://larahack.github.io/linkflows_superpattern/doc/sp/index-en.html}} and in Table \ref{tab:relations}.

The remaining symbols are interpreted as follows. The symbol $i$ stands for the general relation ``is in the context of'' that maps instances in the subject and object classes to instances of the context class.\footnote{A future version of the super-patterns might also make this relation configurable with more specific relations when the super-pattern is instantiated.} $P(\ldots|\ldots)$ represents the conditional probability function of standard statistics, but with the slightly non-standard convention of having logical expressions as elements. They are interpreted as the sets of tuples of all contained free variables (here, $x$ and $y$) that satisfy the expression. $z$, finally, is a bound variable in the first of these expressions with the normal first-order semantics.

We can also phrase the $P(\ldots|\ldots)$ part as a matter of ratios instead of probabilities: It is the ratio of things that satisfy the left-hand side out of all the things that satisfy the right-hand side. This can be defined as such in higher-order logic, but we chose here the notationally simpler expression with conditional probabilities.

To get a better understanding of how the super-pattern works, let us walk through two cases. First, we can have a look at the effect of the \emph{always} qualifier, for which we set ${\lesseqgtr}$ to ${=}$, $q$ to $1$, and $m$ to $\epsilon$. The above expression then becomes (omitting the empty symbol $\epsilon$, as it has no effect by definition):
\begin{multline*}
P( \mbox{~} ( \exists z( o(z) \wedge i(z,x) \wedge r(y,z) ) ) \mbox{~} | \mbox{~} s(y) \wedge c(x) \wedge i(y,x) \mbox{~} ) = 1
\end{multline*}
This stating that the probability is 1, this is equivalent to saying that the right-hand side part of the conditional probability expression implies the left-hand side:
\begin{multline*}
\exists x \exists y ( \mbox{~} s(y) \wedge c(x) \wedge i(y,x) \rightarrow \exists z( o(z) \wedge i(z,x) \wedge r(y,z) ) \mbox{~} )
\end{multline*}
We see that this is expressing that for all things $y$ in class $s$ that are in the context of a thing $x$ in class $c$, there exists another thing $z$ in class $o$ that is in the context of the same thing $x$ such that there is a relation of type $r$ from $y$ to $z$, which matches how we introduced the super-pattern above.

As a second case, let us pick the more complex case of the qualifier \emph{can generally not}, corresponding to the parameter values ${\lesseqgtr} = {\leq}$, $q = 0.1$, and $m = \Diamond$. This gives us:
\begin{multline*}
P( \mbox{~} \Diamond( \exists z( o(z) \wedge i(z,x) \wedge r(y,z) ) ) \mbox{~} | \mbox{~} s(y) \wedge c(x) \wedge i(y,x) \mbox{~} ) \leq 0.1
\end{multline*}
This is stating that given a thing $y$ of type $s$ that is in the context of a thing $x$ of type $c$, the probability (or ``ratio'') that it is possible that there exists a thing $z$ of type $o$ in the context of the same thing $x$ such that there is a relation of type $r$ from $y$ to $z$ is less or equal 0.1. Or in other words, thing $z$ as specified is possible under the given conditions in at most 10\% of the cases.

The ``it is possible that'' here corresponds to the ``can'' in the qualifier and to the logical operator $\Diamond$. It is interpreted according to standard modal logic with possible world semantics with the meaning of ``there is an accessible possible world where the following is true''. As a more intuitive variation, we can also say ``it can be made true that''. As we will see below, we often use these ``can'' qualifiers for statements from applied sciences like computer science or medicine, where findings are often stating that a certain kind of treatment or kind of software \emph{can} solve a particular problem, but only if it is actually applied to it. As an extreme example, there might be strong evidence that a novel kind of medical treatment can treat a certain kind of condition, without a single instance of that actual condition having been treated in this way to that point in time. In that situation we need to express the possibility that it could treat these conditions without necessarily treating them in reality.

For the case when the optional context slot is omitted, we can define a \emph{universal context} class, which contains just one element that by definition has everything in its context. Leaving the context slot empty thereby corresponds to setting it to the universal context. In the expression above, $x$ therefore corresponds to the universal context instance and the $i(\ldots,\ldots)$ components are always true and can be omitted. We therefore arrive at this simpler expression without context:
\[
P( \mbox{~} m( \exists z( o(z) \wedge r(y,z) ) ) \mbox{~} | \mbox{~} s(y) \mbox{~} ) \lesseqgtr q
\]

\subsection{Dataset A} \label{sec:dataset-A}

In order to see if the content of high-level scientific claims from different disciplines can be expressed with our super-pattern, we created a set of randomly selected scientific articles from Semantic Scholar. All data can be found online\footnote{\url{https://github.com/LaraHack/linkflows_claims_dataset}}. The general methodology for creating this Dataset A was as follows: (1) selecting a random sample of articles from different disciplines; (2) identifying a high-level scientific claim from each article; (3) applying the super-pattern on this claim with informal classes; and (4) formalizing the classes by finding existing identifiers or defining new ones.

\paragraph{Step 1: select articles.}
We selected a set of 50 random scientific articles by creating random numbers and then consulting Semantic Scholar\footnote{\url{https://www.semanticscholar.org/}} to retrieve the article with that number as Semantic Scholar corpus ID, if it exists. We then checked whether it matched our inclusion criteria of being written in English, published in 2000 or later, having an abstract, and being a research paper. We repeated this process until we collected 50 such matching articles.

\paragraph{Step 2: identify main claim.}
We check the abstract to find the main scientific claim, finding, or conclusion, e.g. ``In a cross-sectional study in Korean women, obesity showed closest association with knee osteoarthritis when accompanied by metabolic abnormality''. When several candidate claims existed, we chose the most high-level one, or if that did not resolve it, we simply chose the one mentioned first. We discarded articles that failed to clearly mention their main claim in the abstract.
To arrive at a clear understanding of what exactly we want to formalize, we rephrased the found claims as AIDA sentences \citep{Kuhn2013, Kuhn2018}. These are single English sentences that are Atomic, Independent, Declarative, and Absolute. These properties make them a good tool to delineate what should go into the formalization to be created. For the claim above, the corresponding AIDA sentence could be ``Obesity when accompanied by metabolic abnormality is closely associated with knee osteoarthritis''.

\paragraph{Step 3: apply the super-pattern.}
Next, we tried to apply the super-pattern by figuring out what classes would have to go into the three class slots, and which qualifier and relation would be most suitable. At this step, we do not yet check whether the needed classes are defined by any of the existing ontologies, but we simply refer to them by their (likely) class name. For the example given above, we would arrive at something similar to what is shown in Figure \ref{fig:superpattern-diagram}.

\paragraph{Step 4: find identifiers in existing ontologies.}
Next we tried to find identifiers in existing ontologies for the classes needed for the super-pattern instantiation from the last step. For this, we used (1) Wikidata\footnote{\url{https://www.wikidata.org}}, the free knowledge base in which new concepts and properties can be added in a structured way, (2) BioOntology\footnote{\url{https://bioportal.bioontology.org/}}, one of the biggest repositories containing biomedical ontologies and (3) Linked Open Vocabularies (or LOV)\footnote{\url{https://lov.linkeddata.es/}}, a big curated data collection of vocabularies that are reviewed and added continuously. All these chosen sources contain indexed vocabularies and ontologies and as such permit full text searches of concepts and relations.
If we found multiple applicable candidate identifiers, we selected the one that seemed most suitable or best documented.
For the cases where we could not find a matching identifier, we tried to find identifiers for parts of the required concept. These parts could then be used to construct a (partial) class definition, such as ``obesity together with metabolic abnormality'', which can be defined as the intersection of the conditions ``obesity'' and ``metabolic abnormality''.
We mint new identifiers for such complex classes and also for the cases where we could not find any existing identifiers at all.

The process described above was itself the result of an iterative process during which the super-pattern was defined in its current form.

\subsection{Dataset B}
\label{sec:dataset-B}

As the super-pattern was developed based on the claims found during the generation of Dataset A, we created an additional dataset that did not influence the super-pattern and can therefore provide a more reliable picture about what the range of scientific claims the super-pattern is able to express. For this Dataset B, we selected an extra set of 25 claims from another set of 25 randomly selected articles from Semantic Scholar, following the same procedure and criteria as for Dataset A above. We therefore arrived at another 25 formalizations in the form of instantiated super-patterns (to the extent it applied; more on this later).

In contrast to Dataset A, Step 3 for Dataset B was performed several times independently by different knowledge representation experts. Only after the independent super-pattern instances were created, the experts discussed and tried to reach an agreement. This will be explained in more detail below.

\section{Evaluation} 
\label{sec:evaluation}

In this section we will present the design and results of the evaluation for our approach. We performed a descriptive analysis and a vocabulary use analysis for the datasets introduced above. In order to assess how well and how consistently the super-pattern can be applied by knowledge representation experts, we ran a \emph{formalization study} where such experts independently perform super-pattern based formalizations.

\subsection{Descriptive Analysis}

We performed a descriptive analysis on the combined datasets used in this research, dataset A and dataset B, one with 50 claims and the other with 25 other claims respectively.

91\% of the claims (68 out of the total 75 claims) could be expressed with the super-pattern. The remaining ones were rather \emph{too simple} and not too complex to be expressed, as they could be formalized with a simple \emph{subject--predicate--object} triple in RDF style. The super-pattern was therefore successful in covering all the non-trivial scientific claims encountered.
76\% of the claims (57 out of the total 75 claims) used the optional context slot of the super-pattern, whereas the remaining ones 15\% (11 claims) used the simpler version of the pattern without context.

Next we can look at the distribution of qualifiers and relations for both datasets, which is shown in Tables \ref{tab:sp-qualifiers} and \ref{tab:relations}. As we can see from Table \ref{tab:sp-qualifiers}, the most used qualifier is ``generally'' in almost 39\% of cases (29 claims), together with its modal counterpart, ``can generally'' in 23\% of cases (17 claims). The modal negative of qualifiers was never used, while using the negative for qualifiers seems to be less common, in just 6\% of cases (6 claims), while the most used qualifiers are positive with 57\% (43 claims) and modal positive with 25\% (19 claims).
\begin{table}[tbp]
\caption{Usage of qualifiers in the dataset.}
\label{tab:sp-qualifiers}
\begin{center}
\small
\begin{tabular}{c|c@{~}c@{~}c@{~}c@{~}c@{~}}
 qualifier & sometimes & frequently & mostly & generally & always \\ 
 \hline
 positive & 4 & 3 & 3 & 29 & 4 \\ 
 negative &  &  & 1 & 5 &  \\ 
 can positive & 1 &  &  & 17 & 1 \\
 can negative &  &  &  &  &  \\
\end{tabular}
\end{center}
\end{table}
In terms of the relations used in the datasets, Table \ref{tab:relations} shows that relations that express causal relations are the most common with 57\% (43 claims), then the equivalency relation ``is same as'' is the next most used with 16\% (12 claims), then in a smaller ratio, the relations marking numerical comparisons (the ``compares to'' relations) and the relations about spatio-temporal relationships (the ``has spatio-temporal relationship with'') are used in about 6-7\% of cases (7 and 6 claims, respectively).

\begin{table}[tbp]
\caption{Usage of relations in the dataset.}
\label{tab:relations}
\begin{center}
\small
\begin{tabular}{l|r|r}
  & & total \\ 
 relation & count & per group \\
 \hline
  \textbf{is same as} & 12 & 12 \\
 \hline
  \textbf{compares to} & 0 & 7 \\
  has similar value as & 0 \\
  has same value as & 0 \\
  has different value from & 0 \\
  has smaller value than & 1 \\
  has larger value than & 6 \\
 \hline
  \textbf{has causal relationship with} & 0 & 43 \\
  affects & 2 \\
  contributes to & 11 \\
  enables & 6 \\
  inhibits & 2 \\
  prevents & 1 \\
  increases & 4 \\
  decreases & 3 \\
  requires & 4 \\
  causes & 5 \\
  is necessary and sufficient for & 2 \\
  is caused by & 3 \\
 \hline
  \textbf{has spatio-temporal relationship with} & 0 & 6 \\
  includes & 5 \\
  is included in & 0 \\
  co-occurs with & 1 \\
  is followed by & 0 \\
  follows & 0 \\
\end{tabular}
\end{center}
\end{table}

Overall, we see a broad but far from uniform distribution. Certain qualifiers and relation types were much more often used than others, which is in general not very surprising. That most claims are of a positive and causal nature is also consistent with expectations.

For Dataset A we can do an analysis of the classes used from the existing ontologies. We only restrict ourselves here to the top-level classes that could directly be used in the slots of the super-pattern, and we exclude here the classes and relations from existing ontologies that can be used to build a complex class definition.
In terms of the vocabulary usage for the classes for the subject, context, and object slots of the super-pattern, we noticed that most of these classes were not present in existing vocabularies, but had to be newly minted. In only 18 out of the 128 classes (14\%), we could directly use an existing class identifier. Most of the time (16 cases), these identifiers came from Wikidata. Therefore we can say that formalizations that use the super-pattern mostly depend on defining new classes, which can typically only be partially defined from existing concepts and relations.

\subsection{Design of Formalization Study} \label{sec:formalization-study-design}

In order to evaluate the proposed method and our super-pattern, we designed a formalization study where several knowledge representation experts independently apply the super-pattern. The goal of this study was to find out how reliably a super-pattern could be created from a given scientific claim. The extent to which users who are not knowledge representation experts would be able to create such formalizations is beyond the scope of this work.

We have therefore designed a three-stage formalization study where the four co-authors of this article participated as knowledge representation experts. In the first stage, the experts independently instantiate the super-pattern for the given claims. In the second stage, each expert is asked to review all four formalizations (which are anonymized and randomly shuffled) from the first stage and select the best ones. In the last stage, all experts meet, discuss their choices, and are given the opportunity to adjust their selection. All data can be found online\footnote{\url{https://github.com/LaraHack/linkflows_formalization_study}}.

\paragraph{Stage 1.}
The first stage started with an introduction session where the participating knowledge representation experts learned about the details of the super-pattern and could discuss any questions. The formalizations of Dataset A were used as examples for all the experts to have a common and consistent understanding of the super-pattern.
After this joint introduction session, all experts worked independently to apply the super-pattern on the 25 claims of Dataset B. For each of these claims, the participants also had to rate (on a scale from 1 to 5) how confident they were in their formalization. For the class slots, the experts were only asked to provide class names, but not existing class identifiers.

\paragraph{Stage 2.}
In this stage, the participants were asked to independently inspect and review all four formalizations (i.e. the own one as well as the ones from the other participants). These formalizations were randomly shuffled and it was not visible who created which of them. The participants were asked to select the best formalization (in their opinion) and to indicate which formalizations in their view contained clear formalization mistakes. If several equally good formalizations were present, participants were allowed to select more than one ``best'' formalization. After this stage, we can assess how much the participants agreed on the best formalization, and how much they agreed on the presence of formalization mistakes. As the task was complex and multiple valid solutions are possible, we can expect a substantial level of disagreement at this stage.

\paragraph{Stage 3.}
The goal of this last part of the formalization study was to see the extent to which disagreement was resolved when the experts were allowed to directly discuss the cases.
In several discussion sessions, the experts went through all 25 claims and their candidate formalizations. The experts could explain why they chose a particular formalization as the best one and what kind of mistakes they identified. During this discussion, the experts could update their previous decisions, e.g. by choosing a different formalization as the best one, or by recognizing a formalization mistake they previously missed.
After this stage, we can assess the degree of agreement when the experts have the chance to consult the others' opinion and have the chance to react on that. This agreement can however also include the effect of social dynamics, such as participants trying to achieve (or prevent) consensus for social rather than conceptual reasons.

\subsection{Formalization Study Results}
\label{sec:formalization-study-results}

For each of the 25 claims from Dataset B we have four different possible formalizations, therefore 100 formalizations in total.
After Stage 2 (before the joint discussion), 38\% of the individual formalizations were marked as containing a mistake by at least one of the experts. This dropped to just 6\% after the discussion, showing agreement being increased by the discussion. Given the complexity of the task, these mistake ratios seem reasonably low. Overall, just 2\% of the individual formalizations had both, at least one mistake mark as well as at least one best mark (this value didn't change after discussion).

\begin{table*}[tbp]
\begin{center}
\small
\caption{\label{tab:levels-agreement}Levels of agreement in the formalization study.}
\begin{tabular}{l|cc|cc|} 
\multirow{2}{*}{Agreement level: before and after discussion } & 
\multicolumn{2}{c|}{before (Stage 2)} & 
\multicolumn{2}{c|}{after (Stage 3)} \\ \cline{2-5}
& abs. & rel. & abs. & rel.\\
\hline
Level A: Full agreement on best formalization & 2 & 0.08 & 21 & 0.84 \\
Level B: Majority agreement on best formalization & 12 & 0.48 & 3 & 0.12 \\
Level C: Full agreement on absence of formalization mistakes & 9 & 0.36 & 1 & 0.04 \\
Level D: No agreement of the types above & 2 & 0.08 & 0 & 0.00 \\
\end{tabular}
\end{center}
\end{table*}

Table \ref{tab:levels-agreement} summarizes the most important results from the Stages 2 and 3 of the formalization study. In order to analyze the level of agreement between the participants before and after the discussion meeting in Stage 3, we devised an agreement score of four levels A to D. Based on the participants' best and mistake marks given, each of the 25 claims is assigned a level from A (most agreement) to D (least agreement).

In Level A, we include all scientific claims with a formalization that everybody agreed it was the best (or one of several ``best''). In Level B, there is no full agreement but a majority agreement of three out of four participants on the best formalization, with no mistake mark from the remaining participant. Level C includes all scientific claims that are not in levels A or B, but for which there exists at least one formalization for which all participants agree that it contains no mistakes. Level D, finally, applies to all claims that are not in any of the categories above. 

We see in Table \ref{tab:levels-agreement} that full agreement on the best formalization was rare after Stage 2, with only 8\% (2 claims of 25). The slightly less restrictive majority agreement of Level B, however, was achieved in 48\% cases, excluding the perfect agreement cases above. Therefore, in a majority of 56\% of cases full or majority agreement on the best formalization was achieved, which seems quite remarkable given that this happened without discussion. Agreement on a formalization without mistakes (Level C) was achieved in another 36\% of the cases, leaving only 8\% to the lowest level D of no agreement at all. Again, considering the complexity and non-deterministic nature of the task, these seem very favorable outcomes that show that the super-pattern can be applied by knowledge representation experts in a reliable and consistent manner.

Looking at the agreement levels after the discussion meeting (Stage 3), we can observe that for most of the formalizations of the scientific claims, full agreement (84\%) or majority agreement (12\%) is reached. In all cases, there is at least a formalization on which all participants agree that it does not contain any mistakes. As expected, such a discussion session drives up agreement values substantially, which underlines the importance of checking the independent assessments of Stage 2 first. The results show that knowledge representation experts can agree on the best formalization in the vast majority of cases.

\begin{figure}[tbp]
    \centering
    \includegraphics[width=\linewidth]{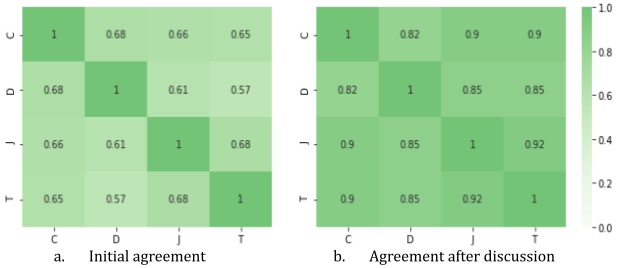}
    \caption{\label{fig:heatmap-agreement}Agreement between participants on the formalizations before and after discussion.}
\end{figure}

The individual pair-wise agreements between the four experts are visualized in Figure \ref{fig:heatmap-agreement}.
To calculate these agreements, we use a value of 1 when the two participants fully agree on the status of an individual formalization (either ``best'', ``mistake'', or no mark), a value of 0.5 when they are different but without conflict (i.e. not ``best'' and ``mistake'' at the same time), and a value of 0 in the case of conflicting ``best'' and ``mistake'' marks. We then take the average of these values of the 100 data points per expert pair (4 formalization candidates for each of 25 claims).

We see that the average values before the discussion (Stage 2) are all above the 0.5 threshold (0.57 being the lowest value), and move quite close to full agreement after the discussion in Stage 3 (all values being over 0.82). Also, these pair-wise disagreement values are quite uniform across the pairs, and disagreement was therefore not driven by just one of the experts. Before as well as after discussion, the uniformity of the disagreement can be seen as an indication that it was rather due to the difficulty of the task and not to fundamental misunderstandings by some of the experts.

These pair-wise results on the individual formalization candidates therefore confirm the more general picture of Table \ref{tab:levels-agreement}. In the future, following the discussion in Stage 3 of the study, minor changes were proposed to the Super-Pattern ontology. As such, a future version might contain small changes to the definitions of some relations where, for example, the definition for the ``includes'' relation, will be ``spatio-temporally or conceptually includes'' instead of just ``spatio-temporally includes''. In summary, we can conclude that writing a formalization that expresses the content of a high-level scientific claim using the super-pattern is an achievable task with a reasonably high success rate. 

\section{Discussion and conclusion}
\label{sec:discussion-conclusion}

While Semantic technologies have been extensively applied to a broad range of applications in various science disciplines, their use is quite limited with regards to science communication. The rapid increase in the number of scientific articles and the outlook to be able to communicate, link, and validate scientific discoveries in an automated manner should make it a priority to allow for the content of scientific articles to become readable not only by humans, but by machines as well. However, the use of these semantic technologies have mainly targeted the structure of the scientific publications themselves and not the content of the core scientific discoveries of these articles. In this research, we managed to show how we can create a framework in which we can represent these scientific claims with formal semantics.

Taking a random sample of scientific claims extracted from scientific articles from all disciplines, we succeeded to create formalizations of these scientific claims by using our super-pattern approach. It appears that this approach works for scientific claims from a variety of disciplines, confirming the cross-disciplinary nature and applicability of the super-pattern. Most of the classes used in the super-pattern formalizations had to be minted and could not be fully defined from existing vocabularies, but we argue that we managed to solve the more difficult part of the problem, as expressing the high-level logical structure of scientific claims is a more complex task than defining new classes. We could also demonstrate that knowledge representation experts can create such formalizations using the super-pattern in a fairly consistent and reliable manner, despite the inherent complexity and difficulty of this task.

In future work, we plan to investigate how our formal super-pattern representations can be used in practice to let researchers themselves, who are not necessarily knowledge representation experts, publish their own scientific claims. Furthermore, it will be interesting to investigate how advanced reasoning can be applied, considering that super-pattern formalizations use higher-order logic with its theoretical and practical problems around efficiency and decidability. In any case, any kind of reasoning with partial results or incomplete heuristics would constitute a huge advance compared to what we can currently do in terms of reasoning on scientific knowledge. We can then imagine countless novel applications, such as overarching aggregations, finding supporting or conflicting claims, answering high-level questions, making visualization of scientific knowledge, and much more. In summary, it would allow us to harness computers in a more effective way to increase our understanding of scientific discoveries and thereby amplify their impact.

\begin{acks}
The authors thank IOS Press and the Netherlands Institute for Sound and Vision, who partly funded this research.
\end{acks}

\bibliographystyle{ACM-Reference-Format}
\bibliography{main.bib}

\end{document}